\documentclass[12pt]{iopart}

\pdfoutput=1

\usepackage{graphicx}
\usepackage{bm}
\usepackage{color}
\usepackage{rotating}

\begin{document}
\title{Parity Detection in Quantum Optical Metrology \\ Without Number Resolving Detectors}

\author{William~N.~Plick$^{1}$, Petr~M.~Anisimov$^{1}$, Jonathan~P.~Dowling$^{1}$, Hwang Lee$^{1}$ and Girish~S.~Agarwal$^{2}$}

\address{$^{1}$Hearne Institute for Theoretical Physics, Department of Physics and Astronomy, Louisiana State University, Baton Rouge, LA 70803.\\
$^{2}$Department of Physics, Oklahoma State University, Stillwater, OK 74078}

\begin{abstract}
\noindent We present a method of directly obtaining the parity of a Gaussian state of light without recourse to photon-number counting. The scheme uses only a simple balanced homodyne technique, and intensity correlation. Thus interferometric schemes utilizing coherent or squeezed light, and parity detection may be practically implemented for an arbitrary photon flux. Specifically we investigate a two-mode, squeezed-light, Mach-Zehnder interferometer and show how the parity of the output state may be obtained. We also show that the detection may be described independent of the parity operator, and that this ``parity-by-proxy'' measurement has the same signal as traditional parity.
\end{abstract}

\maketitle

\section{Introduction}

Quantum metrology, in part, is the investigation of methods of accurate and efficient phase estimation. To this end, various interferometric setups have been proposed. They mainly differ in the light they use and \--- as a consequence of this \--- the detection scheme that is required to obtain the phase information.\cite{SZ}

The most common interferometric setup is to input a strong coherent laser beam into one port of a beam splitter. One output is retained as a reference, while the other is sent out to interact with some phase-bearing process (called the probe beam): this could be a medium, which changes the properties of the light in the presence of magnetic fields \cite{Budker}, to a path difference caused by the presence of a gravitational distortion \cite{russ,weiss}. The change in distance of some distant object could also be measured by reflecting the probe beam off this target. When the light returns it is recombined on a second beam splitter with the reference beam. The amount of light which exits one, or the other, output port of this beam splitter is dependent on the phase difference between the two beams, allowing measurement of the probed object. Typically the intensity of one output port is subtracted from the other, cancelling out phase distortions endemic to the device itself.

There are many variants of this basic machine. One of the most important metrics, by which they are compared, is their ability to measure as small a change in the probe phase as possible. Quantum mechanics puts limits on how sensitive devices can be made. For classical light (that is, coherent light) the limit is known as the shot-noise limit and is given by $1/\sqrt{I}$, where $I$ is the intensity of the inputed light. By taking advantage of quantum ``tricks'' (such as the use of entangled light in the device), the more fundamental Heisenberg Limit may be reached, this limit is given by $1/n$ where \--- in most cases \--- $n$ is the average number of photons in the input. By making adjustments to the measurement scheme the Cramer-Rao bound may be approached, this is a fundamental limit given by information theoretic arguments, and is dependent on the quantum state of the input light.

Parity has been shown to be a very desirous method of detection in interferometry for a wide range of input states. For many input states parity does as well, or nearly as well, as state-specific detection schemes \cite{AL}. Furthermore, as has been reported recently, parity paired with a two-mode, squeezed-vacuum interferometer actually reaches below the Heisenberg limit, achieving the Cramer-Rao bound on phase sensitivity \cite{Petr}. Mathematically, parity detection is described by a simple, single-mode operator

\begin{eqnarray}
\hat{\Pi}=(-1)^{\hat{N}},
\end{eqnarray}

\noindent where $\hat{N}$ is the photon number operator. Hence, parity is simply the evenness or oddness of the photon number in an output mode.

The extreme sensitivity of parity can be explained by examining what happens when the parity operator is back-propagated through a beam splitter (this is the $\hat{\mu}$ operator in the language of Ref. \cite{Petr})

\begin{eqnarray}
\hat{\mu}=\sum_{N=0}^{\infty}\sum_{M=0}^{N}|N-M,M\rangle\langle M,N-M|,
\end{eqnarray}

\noindent where the two positions in the state vectors represent the two modes of the the device. As stated in Ref. \cite{Gao} this operator picks up all the off-diagonal, phase-bearing terms in the density matrix of the two-mode light field, which is the root of its high degree of phase sensitivity.

A potential advantage of parity detection might be metrology in the presence of loss. Many of the states which show the greatest potential phase sensitivity also quickly degrade in lossy environments, limiting their usefulness. In a real-world application of super-sensitive remote sensing it may, or may not, be advantageous to use delicate quantum light \--- depending on the current conditions in the environment. One could imagine an adaptable device which could be used to send one of several different states out into the environment to perform measurement. Perhaps a N00N state \cite{Jon} for low loss, an $M\&M'$ state \cite{Sean} for intermediate loss, and a coherent state for high loss \cite{BMC,other}. The receiver for such a device would not need to be different for each kind of light, but could always use parity detection, adding some robustness and ease of implementation.

Experimentally, parity is often measured, directly, by counting the number of photons in the light field. Unfortunately, number-resolving detectors are difficult to make and operate, and only work at low photon numbers \cite{NR9,waks}. Though there is significant effort and continual progress in this field of quantum optical metrology \cite{NR1,NR2,NR3,NR4,NR5,NR6,NR7,NR8,NR9}, it would be useful to have another way of determining parity. Moreover, photon number counting is more than is necessary to obtain the parity of a state. Alternatively a non-linear optical Kerr interferometer may be used to find the evenness or oddness of a light field \cite{gerry}. But the feasibility of this scheme rests on the availability of materials with large non-linearities. In this paper we present a method of obtaining parity directly \--- without recourse to cumbersome photon number resolving detectors or Kerr materials. Our method is much simpler for states of light, which have Gaussian Wigner functions (a class which includes both squeezed and coherent light). We focus on the two-mode squeezed-vacuum MZI mentioned above and present a complete setup which allows implementation of our promising scheme.

\section{Review of Two-Mode Squeezed Vacuum Fed Mach-Zehnder Interferometry}

\begin{figure}[h]\centering
\includegraphics[scale=0.25]{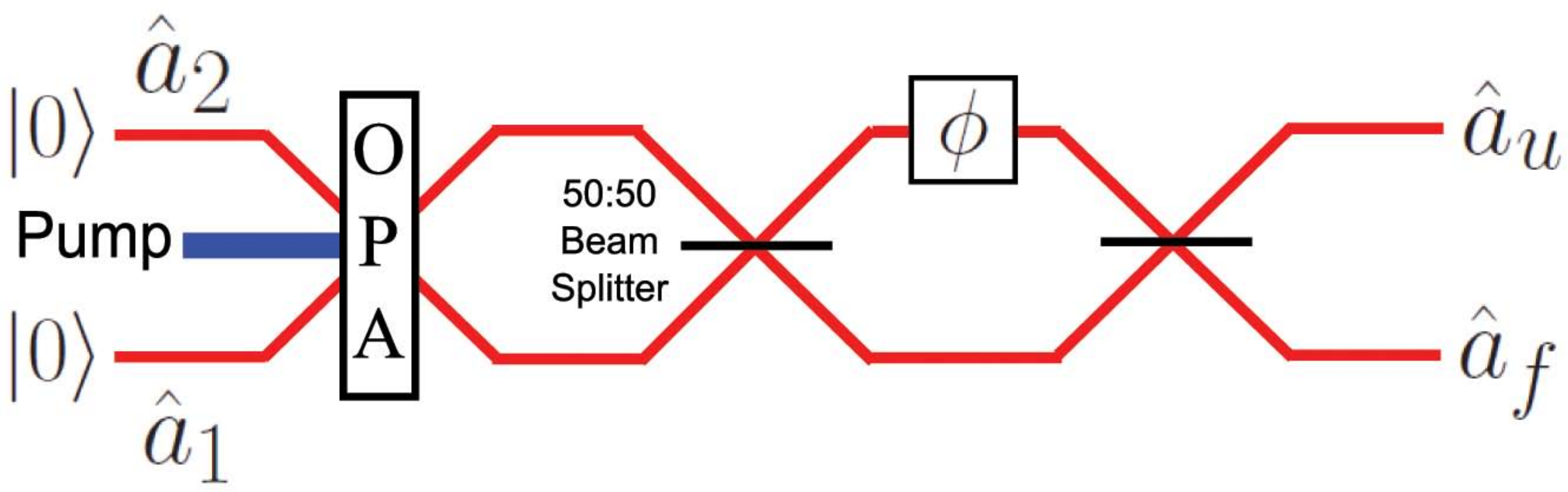}
\caption{\label{TMSVMZI} A diagrammatic representation of a two-mode squeezed vacuum fed MZI. An optical parametric amplifier is pumped with a strong coherent beam, emitting highly correlated photons into two vacuum modes. These modes then interact with a 50:50 beam splitter, a phase shift, and another beam splitter. Parity detection may then be performed on either mode. This setup is both super-sensitive and super-resolving.}
\end{figure}

In a recent paper by Anisimov, et al. \cite{Petr}, it was shown that a MZI fed with a two-mode squeezed vacuum state, and utilizing parity detection, could reach below the Heisenberg limit on phase sensitivity in the limit of few photons. In the high-photon limit the sensitivity goes as $\Delta\phi\simeq 1/\bar{n}$, approximately the Heisenberg limit. For all photon-numbers the setup hits the Cramer-Rao bound. The setup is also super-resolving. A diagram of this kind of setup is provided in Fig. \ref{TMSVMZI}.

Two-mode squeezed vacuum light is produced when a crystal with a high $\chi^{(2)}$ non-linearity is pumped with a laser (there are other methods, but this is the most preferred). Photons are produced which have very high degrees of temporal correlation across two spatial modes. The state vector for this is given by

\begin{eqnarray}
|\xi_{\mathrm{TM}}\rangle&=&\mathrm{exp}\left(re^{-i\Theta}\hat{a}\hat{b}-re^{i\Theta}\hat{a}^{\dagger}\hat{b}^{\dagger}\right)|0,0\rangle\nonumber\\
&=&\frac{1}{\cosh(r)}\sum_{n=0}^{\infty}e^{in\left(\Theta+\pi\right)}[\tanh(r)]^{n}|n,n\rangle,
\end{eqnarray}

\noindent where $r$ is the gain, a parameter characterized by the strength of the non-linearity and the intensity of the pump beam (there are many conventions for defining the gain so it is useful to point out that $r$ is defined completely consistently throughout this paper), $\Theta$ is the phase of the pump beam. The high degree of correlation is clear, photons are produced only pairwise into the two modes. As the light travels through the interferometer the transformation of the operators are described by

\begin{eqnarray}
\left[\begin{array}{c}\hat{a}_{u}\\\hat{a}_{f}\end{array}\right]=\frac{1}{2}\left[\begin{array}{cc}1 & i \\ i & 1\end{array}\right]\left[\begin{array}{cc}1 & 0 \\ 0 & e^{i\phi}\end{array}\right]\left[\begin{array}{cc}1 & i \\ i & 1\end{array}\right]\left[\begin{array}{c}\mu\hat{a}_{1}+\nu\hat{a}_{2}^{\dagger} \\ \mu\hat{a}_{2}+\nu\hat{a}_{1}^{\dagger} \end{array}\right],
\end{eqnarray}

\noindent where $\mu=\cosh{(r)}$ and $\nu=\sinh{(r)}e^{i\Theta}$.

\section{Practical Measurement of Parity of a Single-Mode Field Using Balanced Homodyne}

The relationship between parity of a single-mode field and the value of the Wigner function at the origin has been known for some time \cite{royer}. It is simply

\begin{eqnarray}
\left\langle\hat{\Pi}\right\rangle=\frac{\pi}{2}W(0,0).\label{parity}
\end{eqnarray}

We present an entire and concise derivation of this fact for the sake of clarity and completeness in Appendix A. Wigner functions of unknown quantum states are typically reconstructed after performing optical quantum state tomography \cite{hom}. Homodyne measurements are used to measure the the expectation values of the creation and annihilation operators at different bias phases (each phase constituting a slice). The full Wigner function can then be built up from these images. This would be sufficient to discover the value at the origin and thus the parity, but it is overkill. We only need the value at one point.

Let us consider a specific class of states, called Gaussian states. These states are so called because their Wigner functions are Gaussian in shape. This class includes a very broad category of states, including the coherent states and, conveniently, the squeezed states. It has been shown \cite{ag} that single-mode, Gaussian Wigner functions are given by

\begin{eqnarray}
W(\alpha,\alpha^{*})&=&\frac{1}{\pi\sqrt{\tau^{2}-4|u|^{2}}}e^{-\frac{u(\alpha-\alpha_{o})^{2}+u^{*}(\alpha-\alpha_{o})^{*2}+\tau|\alpha-\alpha_{o}|^{2}}{\tau^{2}-4|u|^{2}}}
\end{eqnarray}

\noindent where $\langle\hat{a}\rangle=\alpha_{o}$, $\langle\hat{a}^{\dagger 2}\rangle-\langle\hat{a}^{\dagger}\rangle^{2}=-2u$, and $\langle\hat{a}^{\dagger}\hat{a}+\frac{1}{2}\rangle-\langle\hat{a}^{\dagger}\rangle\langle\hat{a}\rangle=\tau$. So we need to find $\alpha_{o}$, $u$, $\tau$, and their complex conjugates. Some examples: for coherent states $u=0$, and $\tau=1/2$, for a single-mode squeezed vacuum state $2u=\cosh(r)\sinh(r)e^{-i\Theta}$, and  $\tau=\sinh^{2}(r)+1/2$ (this second equation also holds for a single output port of a TMSV fed MZI, as the output intensities of this device are independent of phase, this is discussed in detail in Appendix B), where $r$ is the gain. Single-mode squeezed vacuum states are defined as

\begin{eqnarray}
|\xi_{\mathrm{SM}}\rangle=e^{\frac{1}{2}\left(re^{-i\Theta}\hat{a}^{2}-re^{i\Theta}\hat{a}^{\dagger 2}\right)}|0\rangle .
\end{eqnarray}

\noindent Also, since squeezing the vacuum state does not change the expectation values of its quadratures $\alpha_{o}=0$ for two mode, and single mode, squeezed vacuum. Hence, for the case of the squeezed vacuum state, we are left with

\begin{eqnarray}
W(0,0)=\frac{1}{\pi\sqrt{\tau^{2}-4|u|^{2}}}.\label{wigner}
\end{eqnarray}

\noindent So now we need only find $\langle\hat{a}^{\dagger}\hat{a}\rangle$, which is simply the intensity of the light. We also need to find $\langle\hat{a}^{\dagger 2}\rangle$ or its complex conjugate, which is the expectation value of a non-Hermitian operator. Thus it can not be measured directly. However the technique of balanced homodyning may be employed to determine it indirectly.

Our analysis is based on the fact that the Gaussian character of the input state does not change after it has gone through the interferometer, as the MZI is a linear device.

\begin{figure}[h]\centering
\includegraphics[scale=0.2]{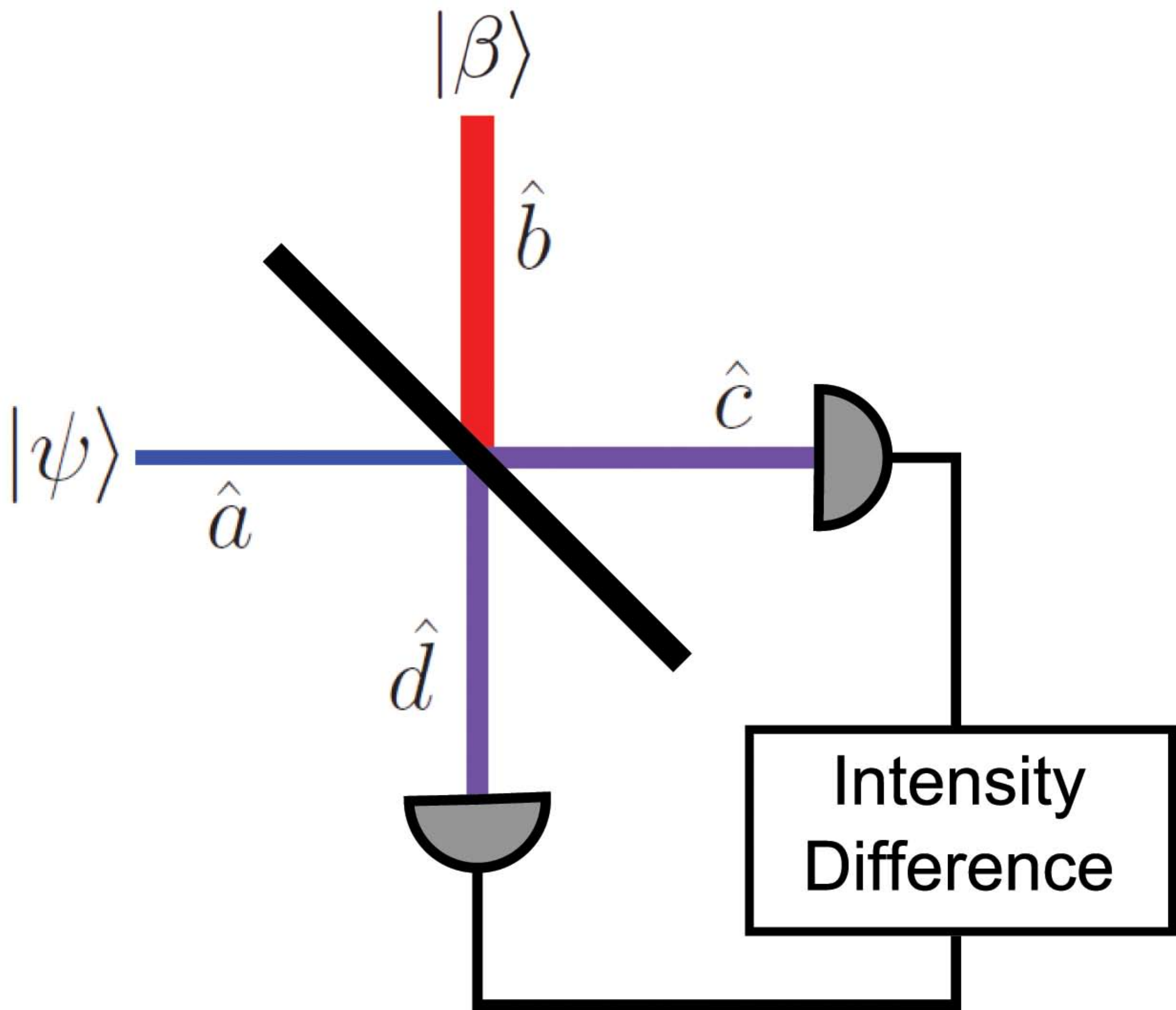}
\caption{\label{hom1}An unknown light beam $|\psi\rangle$ is mixed on a beam splitter with a strong coherent beam of known intensity and phase $|\beta\rangle$, called a local oscillator. Photodetectors are then placed at the output ports. These are connected to a post-processing unit which carries out a intensity differencing. Thus the quadratures of the unknown light may be measured.}
\end{figure}

To begin let's take the standard example of balanced homodyne measurement, originally developed by Yuen and Chan \cite{YC}. The problem that homodyne techniques address is how to measure an, usually inaccessible, quadrature of the light field. The unknown light beam (represented by the state vector $|\psi\rangle$), on which measurements are to be performed, is mixed on a beam splitter with a strong coherent beam of known intensity and phase ($|\beta\rangle$), called a local oscillator. Photo-detectors are then placed at the output ports. See Fig. \ref{hom1}.

\begin{figure}[h]\centering
\includegraphics[scale=0.2]{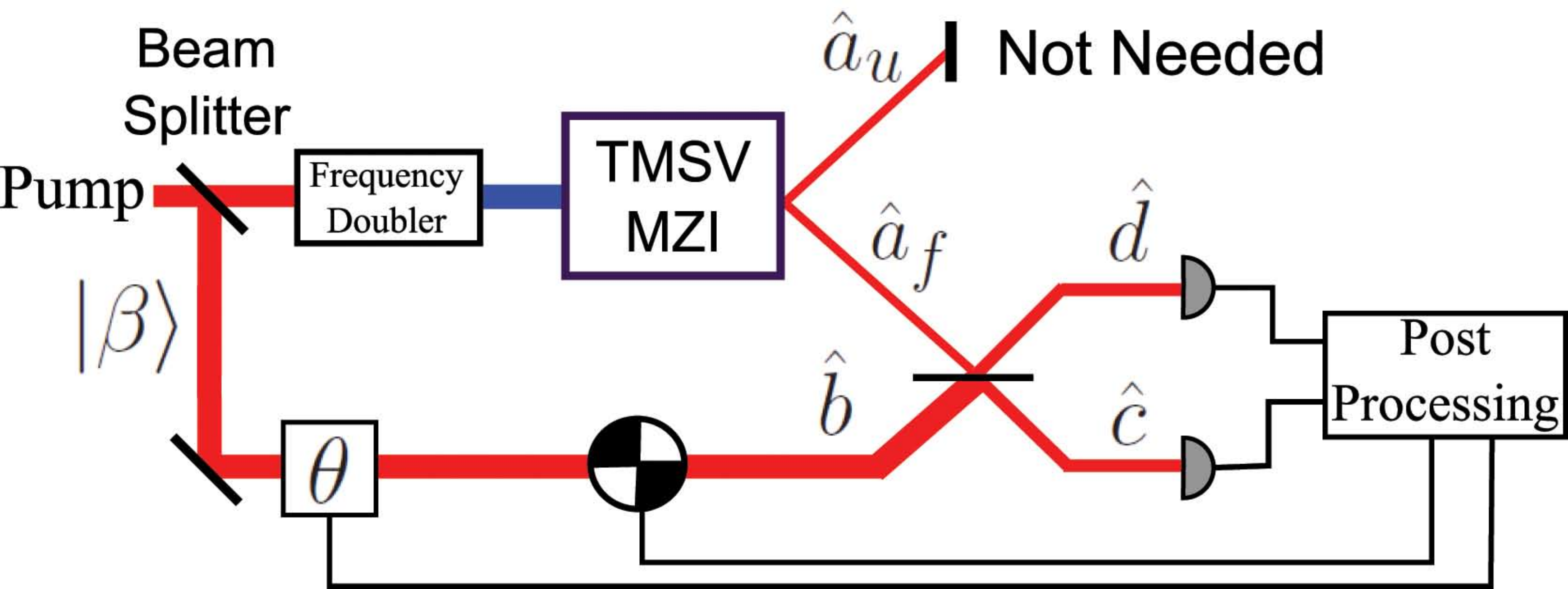}
\caption{\label{hom2} The setup we will discuss: a strong pump beam is sent in from the left. Some of the beam is peeled off for later use as a local oscillator, which can be blocked with a beam chopper or given a different phase with a controllable phase shift. The beam then goes through a frequency doubler before entering the TMSV MZI depicted in Fig. \ref{TMSVMZI}. Thus the squeezed light and the reference beam are of the same frequency, and are phase locked. On output, one mode is mixed on a beam splitter with the local oscillator. After this, two detectors make intensity measurements on the final outputs which are fed into a post processor, which controls the bias phase, and beam chopper placed in the local oscillator beam.}
\end{figure}

For the case of a 50-50 beam splitter we utilize the standard transformations $\hat{c}\rightarrow(\hat{a}+i\hat{b})/\sqrt{2}$ and $\hat{d}\rightarrow(\hat{b}+i\hat{a})/\sqrt{2}$. The intensity difference at the detectors in terms of the input operators is then given by

\begin{eqnarray}
\hat{N}_{D}=\hat{c}^{\dagger}\hat{c}-\hat{d}^{\dagger}\hat{d}=-i(\hat{a}^{\dagger}\hat{b}-\hat{b}^{\dagger}\hat{a}).\label{diff1}
\end{eqnarray}

\noindent The expectation value of Eq. (\ref{diff1}), taken at the input is

\begin{eqnarray}
\langle \psi|\langle\beta |\hat{N}_{D}|\beta\rangle|\psi\rangle=-i|\beta|\left(\langle\hat{a}^{\dagger}\rangle e^{i\eta}-\langle\hat{a}\rangle e^{-i\eta}\right),
\end{eqnarray}

\noindent where $\eta$ is the known phase of the local oscillator. Let's call this expectation value $Y(\eta)$. Now we see that we can measure either $\langle\hat{a}^{\dagger}\rangle$ or $\langle\hat{a}\rangle$ of the unknown light field by making measurements at two different phases of the local oscillator

\begin{eqnarray}
\langle\hat{a}^{\dagger}\rangle&=&\frac{Y\left(\frac{\pi}{2}\right)+iY(0)}{2|\beta|}\nonumber\\
\langle\hat{a}\rangle&=&\frac{Y\left(\frac{\pi}{2}\right)-iY(0)}{2|\beta|}
\end{eqnarray}

Closely related to this is our problem of measuring the second order moments of squeezed light in order to obtain the Wigner function at the origin \--- and thus parity. We will need the second order moment $\langle\hat{a}^{ \dagger 2}\rangle$. To do this take the setup in Fig. \ref{hom2}. A strong pump beam is sent in from the left.  Some of the beam is peeled off for later use as a local oscillator. The beam then goes through a frequency doubler before pumping an OPA. Thus the squeezed light and the reference beam are of the same frequency, and are phase locked. The squeezed light then proceeds through a standard MZI, interacting with the phase to be measured. On output, one mode is mixed on a beam splitter with the local oscillator. After this, two detectors make intensity measurements on the final outputs which are fed into a post processor, which controls the bias phase, and beam chopper placed in the local oscillator beam, according to a prescription which will be discussed shortly. This scheme is designed to measure the second order moments of the lower output port of the MZI.

First  we guess the moment $\hat{d}^{\dagger}\hat{d}\hat{c}^{\dagger}\hat{c}$, which represents the intensity-intensity correlations between the two detectors. When this is  propagated back through the beam splitter it becomes

\begin{eqnarray}
\hat{d}^{\dagger}\hat{d}\hat{c}^{\dagger}\hat{c}\rightarrow\frac{1}{4}\left(\hat{a}_{f}^{\dagger 2}{a}_{f}^{2}+\hat{b}^{2}\hat{a}_{f}^{\dagger 2}+\hat{b}^{\dagger 2}\hat{a}_{f}^{2}+\hat{b}^{\dagger 2}\hat{b}^{2}\right).
\end{eqnarray}

\noindent Taking the expectation value, we define the function

\begin{eqnarray}
X(\theta, |\beta|)&\equiv&4\langle\hat{d}^{\dagger}\hat{d}\hat{c}^{\dagger}\hat{c}\rangle\nonumber\\
&=&\langle\hat{a}_{f}^{\dagger 2}\hat{a}_{f}^{2}\rangle+|\beta|^{2}e^{2i\theta}\langle\hat{a}_{f}^{\dagger 2}\rangle+|\beta|^{2}e^{-2i\theta}\langle\hat{a}_{f}^{2}\rangle+|\beta|^{4},
\end{eqnarray}

\noindent which constitutes a single measurement. It is $\langle\hat{a}_{f}^{\dagger 2}\rangle$ that we wish to obtain. We can accomplish this by performing three $X$ measurements at, $\theta=0$, $\theta=\pi/4$, and $|\beta|=0$ (i.e. when the beam is blocked with the beam chopper), and arranging them according to the prescription

\begin{eqnarray}
\langle\hat{a}_{f}^{\dagger 2}\rangle&=&\frac{1}{2i|\beta|^{2}}\left[iX(0,|\beta|)+X(\pi/4,|\beta|)-(i+1)X(0,0)-(i+1)|\beta|^{4}\right]\label{a}
\end{eqnarray}

\noindent We can obtain $\langle\hat{a}_{f}^{\dagger}\hat{a}_{f}\rangle$ easily with $\langle\hat{d}^{\dagger}\hat{d}\rangle+\langle\hat{c}^{\dagger}\hat{c}\rangle-|\beta|^{2}$. With this information and Eq. (\ref{wigner}) we can obtain the parity of a TMSV fed MZI. It should be noted that we require three measurements in order to reconstruct parity. It is assumed that the phase varies on a time scale which is slower than the speed of the optical elements performing the measurements.

But, does this detection scheme really produce the same signal as parity? To answer this question we can remove any mention of parity from the calculation and directly compute the signal of the detection protocol. In order to do this we employ an operator propagation technique. The operators at the output are related to the operators at the input by the matrix transformations in Eq (\ref{trans}). The first matrix (from right to left) represents the Bogoliubov transformation of the OPA, where $\mu=\cosh(r)$, $\nu=\sinh(r)$, and $r$ is the gain. The phase of the pump has been set to zero. The next four matrices represent the first beam splitter, the probe and control phases, the second beam splitter, and the homodyning beam splitter. We then can write down the the output operators, which we use during detection, in terms of the input operators, where taking the expectation values is more tractable. Despite this a specially written computer code is still required to compute these expectation values.

\begin{tiny}
\begin{eqnarray}
\left[\begin{array}{c}
\hat{a}_{u}\\ \hat{a}_{u}^{\dagger}\\ \hat{d}\\ \hat{d}^{\dagger}\\ \hat{c}\\ \hat{c}^{\dagger}
\end{array}\right]&=&\frac{1}{2\sqrt{2}}\left[\begin{array}{cccccc}
\sqrt{2}&0&0&0&0&0 \\ 0&\sqrt{2}&0&0&0&0 \\ 0&0&1&0&i&0 \\ 0&0&0&1&0&-i \\ 0&0&i&0&1&0 \\ 0&0&0&-i&0&1
\end{array}\right]
\left[\begin{array}{cccccc}
1&0&0&0&i&0 \\ 0&1&0&0&0&-i \\ 0&0&\sqrt{2}&0&0&0 \\ 0&0&0&\sqrt{2}&0&0 \\ i&0&0&0&1&0 \\ 0&-i&0&0&0&1
\end{array}\right]\nonumber\\
& &\times
\left[\begin{array}{cccccc}
e^{i\phi}&0&0&0&0&0 \\ 0&e^{-i\phi}&0&0&0&0 \\ 0&0&e^{i\theta}&0&0&0 \\ 0&0&0&e^{-i\theta}&0&0 \\ 0&0&0&0&1&0 \\ 0&0&0&0&0&1
\end{array}\right]\left[\begin{array}{cccccc}
1&0&0&0&i&0 \\ 0&1&0&0&0&-i \\ 0&0&\sqrt{2}&0&0&0 \\ 0&0&0&\sqrt{2}&0&0 \\ i&0&0&0&1&0 \\ 0&-i&0&0&0&1
\end{array}\right]\nonumber\\
& &\times\left[\begin{array}{cccccc}
\mu&0&0&0&0&\nu \\ 0&\mu&0&0&\nu&0 \\ 0&0&1&0&0&0 \\ 0&0&0&1&0&0 \\ 0&\nu&0&0&\mu&0 \\ \nu&0&0&0&0&\mu
\end{array}\right]
\left[\begin{array}{c}
\hat{a}_{1} \\ \hat{a}_{1}^{\dagger} \\ \hat{b} \\ \hat{b}^{\dagger} \\ \hat{a}_{2} \\ \hat{a}_{2}^{\dagger}
\end{array}\right]\nonumber
\end{eqnarray}
\end{tiny}

\vspace{-0.8in}

\begin{eqnarray}
\label{trans}
\end{eqnarray}

A single measurement of $X$ evaluates to

\begin{eqnarray}
X(\theta,|\beta|)&=&\frac{1}{16}\left[11+16\beta^{4}+\cos(2\phi)-16\cosh(2r)\right.\nonumber\\
& &\left.+16\beta^{2}(\cos(2\theta-\phi)\sin(\phi)\sinh(2r)\right.\nonumber\\
& &\left.-(\cos(2\phi)-5)\cosh(4r)\right].
\end{eqnarray}

\noindent Performing the three measurements and using Eq. (\ref{a}) we obtain

\begin{eqnarray}
\langle\hat{a}_{f}^{\dagger 2}\rangle=e^{-i\phi}\cosh(r)\sinh(r)\sin(\phi).\label{a2}
\end{eqnarray}

\noindent For the sake of simplicity we take the device to be lossless, making $\langle\hat{n}_{f}\rangle=\sinh^{2}(r)$, as the intensity output of a TMSV fed MZI is independent of phase, that is

\begin{eqnarray}
\langle\mathrm{out}|\hat{a}_{f}^{\dagger}\hat{a}_{f}|\mathrm{out}\rangle=\langle\mathrm{out}|\hat{a}_{u}^{\dagger}\hat{a}_{u}|\mathrm{out}\rangle=\sinh^{2}(r).
\end{eqnarray}

\noindent There is a quick and intuitive proof of this somewhat surprising outcome in Appendix B.

Using Eq. (\ref{wigner}) we can write the signal of the detection protocol, $S$, in terms of these operators as

\begin{eqnarray}
S=\frac{1}{2\sqrt{\left(\langle\hat{n}_{f}\rangle+\frac{1}{2}\right)^{2}-\left|\langle\hat{a}_{f}^{\dagger 2}\rangle\right|^{2}}}.
\end{eqnarray}

\noindent Substituting our calculated expectation values, Eq. (\ref{a2}), and making the choice of bias phase $\phi\rightarrow\phi+\pi/2$ this expression becomes

\begin{eqnarray}
S=\frac{1}{\sqrt{1+\bar{n}(\bar{n}+2)\sin^{2}(\phi)}}.
\end{eqnarray}

\noindent Which is identical to the signal for parity detection given in Ref. \cite{Petr}, where $\bar{n}$ is the total average number of photons exiting the OPA and is equal to $2\langle\hat{n}_{f}\rangle$. Thus we conclude that our homodyne technique reproduces exactly parity detection. Though we have not computed explicitly the quantum noise for this setup, it is reasonable to assume that since the signal is precisely the same the noise will be equivalent. The calculated minimum detectable phase shift for the Anisimov, et al., setup is given in Ref. \cite{Petr} as

\begin{eqnarray}
\Delta\phi_{\mathrm{min}}\simeq\frac{1}{\sqrt{\bar{n}(\bar{n}+2)}},
\end{eqnarray}

\noindent in the vicinity of $\phi=0$. It needs to be pointed out that our scheme requires three to four interrogations of the light field whereas a ``true'' parity measurement would only require one. Thus, in principle, the $\bar{n}$ in our scheme's sensitivity equation should be multiplied by a factor of three to four. However, in practice our scheme relies on very well developed detector technologies with short dead times, any potential alternative technology which might be developed to determine parity would likely not have the advantage of these quick detection times, thus it is likely that despite the additional measurements our scheme would be comparatively advantageous. Furthermore it is worth making explicit that the closest competitor to our scheme \--- a full quantum-tomographic reading of the light field \--- would require hundreds of measurements to complete.

\section{Alternate Setup}

\begin{figure}[h]
\includegraphics[scale=0.2]{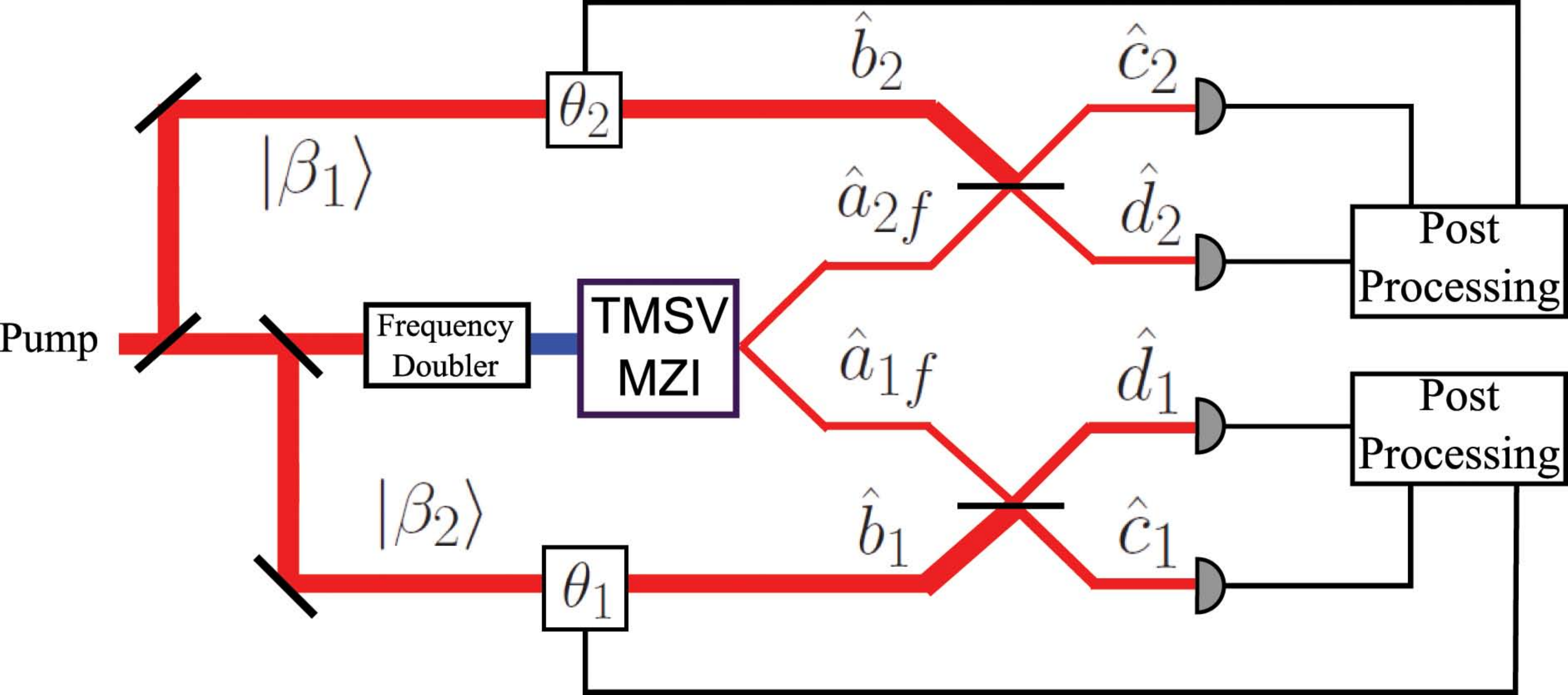}
\caption{\label{altset} The alternate setup: a strong pump beam is sent in from the left. Two beam splitters peel of some of the pump beam for use later as local oscillators for each output port. On output, both modes are mixed on a beam splitter with local oscillators. After this four detectors make intensity measurements which are fed into post processors which control the bias phases placed in the local oscillator beams and computes parity in both ports. Also, note that the beam choppers have been removed.}
\end{figure}

We would also like to present an alternate setup for obtaining the parity of a TMSV MZI. This setup is presented in Fig. \ref{altset}.

Here, instead of performing parity by proxy on only one port we perform it on both simultaneously. Since we are performing intensity measurements across two output modes, fluctuations which are not due to changes in phase may be subtracted out. This will allow noise, which is endemic to the device, to be compensated for. Furthermore this may provide some robustness to loss, as we are now no longer ignoring half the light outputted from the device.

Also, note that the beam choppers have been removed, this is because this device also demonstrates an alternate way of eliminating the unwanted terms using only the controlled phase shift. Not using the beam chopper has the advantage of keeping a more consistent intensity on the photodetectors, which may allow for more sensitive devices to be used. The price of their removal however is that four measurements at four different settings of the bias phase are required (as opposed to only three in the previous setup).

It should be noted that the removal of the beam chopper is not related to the second set of photodetectors. This setup demonstrates two conceptually separate modifications. They are presented together because they are both likely to be experimentally advantageous, if conceptually more complicated than our original case.

Let us take the bottom port, and redefine the $X$ measurement as

\begin{eqnarray}
X_{1}'(\theta_{1})&\equiv&4\langle\hat{d}_{1}^{\dagger}\hat{d}_{1}\hat{c}_{1}^{\dagger}\hat{c}_{1}\rangle\nonumber\\
&=&\langle\hat{a}_{1f}^{\dagger 2}\hat{a}_{1f}^{2}\rangle+|\beta_{1}|^{2}e^{2i\theta}\langle\hat{a}_{1f}^{\dagger 2}\rangle+|\beta_{1}|^{2}e^{-2i\theta}\langle\hat{a}_{1f}^{2}\rangle+|\beta_{1}|^{4},
\end{eqnarray}

\noindent where we are now no longer concerned with the intensity of the local oscillator. We can obtain the desired moment by performing this measurement four times, at four different phases, according to the prescription

\begin{eqnarray}
\langle\hat{a}_{1f}^{\dagger 2}\rangle=\frac{iX_{1}'(0)+X_{1}'\left(\frac{\pi}{4}\right)-iX_{1}'\left(\frac{\pi}{2}\right)-X_{1}'\left(-\frac{\pi}{4}\right)}{4i|\beta_{1}|^{2}}.
\end{eqnarray}

\noindent Again this can be used to obtain parity, likewise for the upper port. Note that the measurement of parity on either the upper or lower port does not involve the other port, allowing two independent readings of the change in phase, adding a built in redundancy.

\section{Summary}

To conclude we have devised a method by which the very desirable parity measurement may be performed on Gaussian states. We have shown in detail how to set up a parity detector for the specific case of squeezed light using homodyning, thus realizing the possibility of practical sub-Heisenberg phase estimation in a Mach-Zehnder Interferometer. We also showed how, though the detection scheme mimics parity measurement, it may be considered conceptually independent and achieves exactly the same signal as parity.

It is also useful to point out that the parity of Non-Gaussian fields may be obtained via Eq. (\ref{parity}), however in this case a measurement of the full Wigner distribution, using quantum state tomography, is required.

\vspace{0.3in}

\section*{Acknowledgements}

W.N.P. would like to acknowledge the Louisiana Board of Regents and The Department of Energy for funding. In addition J.P.D. and P.M.A. acknowledge support from the Army Research Office, the Boeing Corporation, the Foundational Questions Institute, the Intelligence Advanced Research Projects Activity, and the Northrop-Grumman Corporation.

\section*{Appendix A}

First, start with the characteristic function of the Wigner function $C_{W}=\mathrm{Tr}[\hat{\rho}\hat{D}(\lambda)]$, where $\hat{\rho}$ is the density matrix, and $\hat{D}(\lambda)$ is the displacement operator. Accordingly the Wigner function is defined by

\begin{eqnarray}
W(\alpha,\alpha^{*})=\frac{1}{\pi^{2}}\int d^{2}\lambda e^{\lambda^{*}\alpha-\lambda\alpha^{*}}C_{W}(\lambda),
\end{eqnarray}

\noindent which at the origin then becomes

\begin{eqnarray}
W(0,0)=\frac{1}{\pi^{2}}\int d^{2}\lambda\mathrm{Tr}\left[e^{\lambda\hat{a}^{\dagger}-\lambda^{*}\hat{a}}\hat{\rho}\right].
\end{eqnarray}

\noindent Now using the Cambell-Baker-Hausdorf (CBH) theorem, $e^{\hat{A}+\hat{B}}=e^{-\frac{1}{2}[\hat{A},\hat{B}]}e^{\hat{A}}e^{\hat{B}}$ (as long as $\hat{A}$ and $\hat{B}$ commute with their commutator), we rewrite

\begin{eqnarray}
W(0,0)=\frac{1}{\pi^{2}}\int d^{2}\lambda e^{-\frac{1}{2}|\lambda |^{2}}\langle e^{\lambda\hat{a}^{\dagger}}e^{-\lambda^{*}\hat{a}}\rangle,
\end{eqnarray}

\noindent keeping in mind that the trace over the density matrix tensored with an operator is the expectation value of that operator.  Expanding out the exponentials obtains

\begin{eqnarray}
W(0,0)&=&\frac{1}{\pi^{2}}\int d^{2}\lambda e^{-\frac{1}{2}|\lambda |^{2}}\left\langle\sum_{q=0}^{\infty}\frac{(\lambda\hat{a}^{\dagger})^{q}}{q!}\sum_{p=0}^{\infty}\frac{(-\lambda^{*}\hat{a})^{p}}{p!}\right\rangle\nonumber\\
&=&\int^{\infty}_{0}\int^{2\pi}_{0}d|\lambda |d\theta |\lambda |\frac{e^{-\frac{1}{2}|\lambda |^{2}}}{\pi^{2}}\left\langle\sum_{p=q}\frac{(-|\lambda |^{2})^{p}\hat{a}^{\dagger p}\hat{a}^{p}}{(p!)^{2}}\right.\nonumber\\
& &\left.+\sum_{p\neq q}\frac{(-1)^{p}|\lambda |^{p+q}e^{i\theta(q-p)}\hat{a}^{\dagger q}\hat{a}^{p}}{p!q!}\right\rangle.
\end{eqnarray}

\noindent In the second line we have switched to polar coordinates ($\lambda = |\lambda |e^{i\theta}$) and rearranged the sums into terms where $p=q$ and terms where $p\neq q$. Now note that

\begin{eqnarray}
\int^{2\pi}_{0}d\theta e^{i\theta x}=0,
\end{eqnarray}

\noindent for all integers $x$. So all the terms in the second sum integrate to zero and we have, after some rearranging,

\begin{eqnarray}
W(0,0)&=&\frac{1}{\pi^{2}}\sum_{p}\int^{\infty}_{0}\int^{2\pi}_{0}d|\lambda |d\theta |\lambda |\nonumber\\
& &\times e^{-\frac{1}{2}|\lambda |^{2}}(-|\lambda |^{2})^{p}\left\langle\frac{\hat{a}^{\dagger p}\hat{a}^{p}}{(p!)^{2}}\right\rangle.
\end{eqnarray}

\noindent The integral over $|\lambda |$ is not trivial but it can be changed to a known form with some substitution, the integral over $\theta $ is straightforward, leaving

\begin{eqnarray}
W(0,0)&=&\frac{1}{\pi}\sum_{p}\frac{(-1)^{p}2^{p+1}}{p!}\left\langle\frac{\hat{a}^{\dagger p}\hat{a}^{p}}{(p!)^{2}}\right\rangle\nonumber\\
&=&\frac{2}{\pi}\sum_{n=0}^{\infty}C_{n}\langle n|\sum_{p=0}^{n}\frac{(-2)^{p}}{p!}\hat{a}^{\dagger p}\hat{a}^{p}|n\rangle,
\end{eqnarray}

\noindent where in the second line the expectation value is taken to be explicitly in the number basis (note that the sum over {p} can now only extend to $n$). The operators acting on the right number state produce

\begin{eqnarray}
W(0,0)=\frac{2}{\pi}\sum_{n=0}^{\infty}C_{n}\sum_{p=0}^{n}\left(\begin{array}{c}n\\p\end{array}\right)(1)^{n}(-2)^{p},
\end{eqnarray}

\noindent where all the factorials have been written as $n$-choose-$p$. Also note that we have multiplied by one. The sum over $p$ is now clearly the binomial expansion of $(1-2)^{n}$. And so all that remains is

\begin{eqnarray}
W(0,0)=\frac{2}{\pi}\sum_{n=0}^{\infty}C_{n}(-1)^{n}=\frac{2}{\pi}\left\langle(-1)^{\hat{N}}\right\rangle.
\end{eqnarray}

\noindent The expectation value of the parity operator times a constant. Q.E.D.

\section*{Appendix B}

We now show why the intensity of a TMSV MZI is independent of phase. This effect is due to the fact that when a two-mode squeezed vacuum state, $|\xi_{\mathrm{TM}}\rangle$, is incident on both ports of a 50:50 beam splitter, for example in Fig. \ref{TMSVMZI}, the output is the product of two single-mode squeezed vacuum states, $|\xi_{C}\rangle|\xi_{D}\rangle$, where $C$ and $D$ label the modes. Thus, for beam splitter, and phase shift transformations: $\hat{B}$, and $\hat{P}$, the intensity of output mode $f$ can be written as

\begin{eqnarray}
I_{f}&=&\langle\xi_{\mathrm{TM}}|\hat{B}^{\dagger}\hat{P}^{\dagger}_{C}\hat{B}^{\dagger}\hat{a}_{f}^{\dagger}\hat{a}_{f}\hat{B}\hat{P}_{C}\hat{B}|\xi_{\mathrm{TM}}\rangle\nonumber\\
&=&\langle\xi_{C}|\langle\xi_{D}|\hat{P}^{\dagger}_{C}\hat{B}^{\dagger}\hat{a}_{f}^{\dagger}\hat{B}\hat{B}^{\dagger}\hat{a}_{f}\hat{B}\hat{P}_{C}|\xi_{C}\rangle|\xi_{D}\rangle\nonumber\\
&=&\langle\xi_{C}|\langle\xi_{D}|\hat{P}^{\dagger}_{C}(\hat{C}^{\dagger}-i\hat{D}^{\dagger})(\hat{C}+i\hat{D})\hat{P}_{C}|\xi_{C}\rangle|\xi_{D}\rangle\nonumber\\
&=&\langle\xi_{C}|\langle\xi_{D}|\hat{C}^{\dagger}\hat{C}+\hat{D}^{\dagger}\hat{D}-i\hat{D}^{\dagger}\hat{C}e^{i\phi}+i\hat{C}^{\dagger}\hat{D}e^{-i\phi}|\xi_{C}\rangle|\xi_{D}\rangle.
\end{eqnarray}

\noindent The operators $\hat{C}$ and $\hat{D}$ are the mode operators of the the two inputs to the final MZI beam splitter. Only the last two terms carry phase information and, since for squeezed vacuum states the expectation values of first order moments is zero, the phase dependence is eliminated.

\end{document}